\documentclass[twocolumn,showpacs,preprintnumbers,amsmath,amssymb,floatfix,superscriptaddress,aps]{revtex4}

\usepackage{graphicx}% Include figure files
\usepackage{dcolumn}% Align table columns on decimal point
\usepackage{bm}% bold math

\newcommand{\be}{\begin{equation}}
\newcommand{\ee}{\end{equation}}
\newcommand{\beq}{\begin{eqnarray}}
\newcommand{\enq}{\end{eqnarray}}
\newcommand{\ua}{\uparrow}
\newcommand{\da}{\downarrow}

\begin{document}

%\preprint{APS/123-QED}

\title{Density and spin response function of a normal Fermi gas at unitarity}  

%\date{\today}

\author{S. Stringari}
\affiliation{Dipartimento di Fisica, Universit\`a di Trento and CNR-INFM BEC Center, I-38050 Povo, Trento, Italy}

\begin{abstract}
Using Landau theory of Fermi liquids we calculate the dynamic response of both a polarized and unpolarized  normal Fermi gas at zero temperature in the strongly interacting regime of large scattering length. We show that at small excitation energies the {\it in phase} (density) response is enhanced with respect to the ideal gas prediction  due to the increased compressibility. Viceversa, the {\it out of phase} (spin) response is quenched as a consequence of the tendency of the system to pair opposite spins. The long wavelength behavior of the static structure factor is explicitly calculated. The results  are compared with the predictions in the collisional and superfluid  regimes. The emergence of a spin zero sound solution in the unpolarized normal phase is explicitly discussed. 

\end{abstract}

%\pacs{Valid PACS appear here}

\maketitle

There is now  clear experimental evidence that, in the unitary limit of  large  scattering lengths, an harmonically trapped  polarized   Fermi gas  
 phase separates, at very low temperature, into a superfluid core sorrounded by a  polarized normal component \cite{MIT1,Rice,MIT2}. When the total polarization of the gas exceeds a critical value the superfluid component disappears and the gas becomes normal.   The observed phase separation and the  shape of the density profiles  in large atomic samples is theoretically well understood employing the equation of state of the superfluid and normal phases in the  local density approximation \cite{rmp,lobo}. These studies have revealed  the crucial role played by the interactions in the normal phase.  The maximum local concentration (Chandrasekhar-Clogston limit) of the minority (spin down) versus the majority (spin up) component achieved at unitarity is predicted to be  $x=n_{\downarrow}/n_{\uparrow} \sim 0.45$ at zero temperature, in agreement with the most recent experiments carried out at MIT \cite{MIT2}. Larger concentrations are expected to occur if one imposes  an adiabatic  rotation to the gas which favours the formation of the normal phase \cite{ias}. Actually in the rotating case  even an unpolarized Fermi gas is expected to give rise, at unitarity, to phase separatation. 
 
 The availability of the  normal phase of the unitary Fermi gas at very low temperatures opens new stimulating perspectives, due to the expected Fermi liquid behavior of   this strongly interacting system.  Previous studies of  the  dynamic behavior have mainly focused on the particle (spectral) response (see, for example, \cite{levin} and references therein), for which relevant information is experimentally available through radio-frequency transitions \cite{rfexp}. A quantitative comparison between theory and experiment is however still an open issue, especially due to the non trivial role played by   final state interactions. The motion  of single impurities in harmonically trapped configurations has also been  the object of  theoretical investigation \cite{lobo}.
 
 The purpose of this letter is to study the density and spin density dynamic response function of the polarized normal phase. With respect to the motion of single impurities, the excitations investigated in the present work involve wavelengths much smaller than the size of the gas and they consequently correspond to more favourable conditions for reaching the collisionless regime where the most interesting features of the Fermi liquid behavior show up. Experimentally the dynamic  response can be measured via two-photon Bragg spectroscopy, a technique already successfully applied to Bose-Einstein condensed gases \cite{kurn} (first Bragg spectroscopic measurements in ultracold Fermi gases  have been recently carried out at high momentum transfer \cite{australia}). Differently from {\it rf} transitions Bragg experiments do not change the internal atomic states and consequently final state effects are absent. Furthermore, by a proper choice of the detuning of the laser beams with respect to the atomic resonance one can measure suitable combinations of the density and spin responses. Finally, using  focused laser beams, one can in principle measure the   response locally, thereby providing valuable information on the uniform matter behavior. Theoretically the density and spin responses of the normal phase can be calculated using Landau 
theory of Fermi liquids \cite{pines}.  This theory has been  so far  mainly applied to unpolarized samples like liquid He3. It has been also developed to study spin transverse excitations in slightly polarized samples \cite{spinpolarized,bedell}. In this work we consider arbitrary polarized Fermi gases.

The basic assumption of Landau's theory is that the system can be described in terms of a gas of long-living quasi-particles interacting through a mean field. This assumption is usually guaranteed at sufficiently low temperature for excitations  close to the Fermi surface of each spin component.  Under these assumptions collisions between quasi-particles can be ignored. A key feature of the theory is the occurrence of deformations of the Fermi surface 
which deeply distinguish  the dynamic behavior of a normal liquid from the one of a superfluid.  In some cases these deformations  give rise to a new type of collective motion, the so-called zero sound.  For sake of simplicity in the following we will  take into account only the interaction terms arising from the isotropic deformations of the Fermi surface, i.e. from the density modulations of the two spin components. In particular we will ignore  effective mass effects which are predicted to be small in the polarized phase at unitarity \cite{lobo,comb,prok,seb}. 
According to the above assumption the relevant interaction terms  can be derived from the interaction energy functional
\begin{equation}
E_{int}=\int d{\bf r}\ e_{int}(n_{\ua},n_{\da}) \; .
\label{Eint}
\end{equation}
The single quasi-particle (qp) Hamiltonian for each $\sigma$-spin component ($\sigma = \uparrow, \downarrow$) will then contain an interaction contribution easily derivable from Eq.(\ref{Eint}): 
$H^{(qp)}_{\sigma}= H_0 + \partial e_{int} / \partial n_\sigma$
where $H_0$ is the free particle Hamiltonian. The quasi-particle Hamiltonian will be modified during the  motion by  position and time dependent terms, associated with the local changes in the density distribution of the two spin species. These terms have   to be determined with a self-consistent procedure. In the linear approximation one has
\begin{equation}
H^{(qp)}_{\sigma}= H_0 + \left({\partial e_{int} \over \partial n_\sigma}\right)_0+\left({\partial^2 e_{int} \over \partial n_\sigma \partial n_\ua}\right)_0 \delta n_\ua  + \left({\partial^2 e_{int} \over \partial n_\sigma \partial n_\da}\right)_0\delta n_\da
\label{deltaHqp}
\end{equation}
where the suffix $0$ indicates that the derivatives should be calculated at equilibrium and $\delta n_\sigma \equiv \delta n_\sigma({\bf r},t)$ 
are the space and time dependent density changes with respect to equilibrium.
The dynamic response function is calculated by adding an external field of the form $V_{ext}=\alpha_\sigma e^{i({\bf q \cdot r}-\omega t)}$ to the Hamiltonian (\ref{deltaHqp}) and the linear  response  $\chi_{\sigma,\sigma^\prime}$ is defined by the induced density fluctuations according to
\begin{equation}
\delta n_\sigma({\bf r},t) = \sum_{\sigma^\prime}\alpha_{\sigma^\prime}\chi_{\sigma,\sigma^{\prime}}(q,\omega) e^{i({\bf q \cdot r}-\omega t)} \; .
\label{chi}
\end{equation}
By inserting result (\ref{chi})  into Eq.(\ref{deltaHqp}) we can calculate the  response of the system using the free particle Hamiltonian by simply replacing  the external coupling $\alpha_\sigma$  with the effective  value
\begin{equation}
\alpha^{eff}_\sigma = \alpha_\sigma + \sum_{\sigma^{\prime}}\alpha_{\sigma^\prime}\left({\partial^2 e_{int} \over \partial n_\sigma \partial n_{\sigma^{\prime}}}\right)_0 \chi_{\sigma,\sigma^{\prime}} 
\label{lambdaeff}
\end{equation}
in $V_{ext}$. This procedure yields a self-consistency relationship for the dynamic response functions and provides the relevant expressions for $\chi_{\sigma,\sigma^\prime}$ in terms of the free particle responses 
$\chi^{0}_{\ua,\ua}$ and $\chi^{0}_{\da,\da}$ (the crossed free response $\chi^{0}_{\ua,\da}$ identically vanishes because  there are no correlations between spin-up and spin-down particles in the ideal gas):
 \beq
&\chi_{\ua,\ua}(q,\omega)&=\chi^0_{\ua,\ua}(q,\omega) \left[1-\chi^0_{\da,\da}(q,\omega){\partial^2 e_{int} \over \partial n_\da \partial n_\da}\right] /D(q,\omega)\nonumber\\
&\chi_{\da,\da}(q,\omega)&=\chi^0_{\da,\da}(q,\omega)\left[1-\chi^0_{\ua,\ua}(q,\omega){\partial^2 e_{int} \over \partial n_\ua \partial n_\ua}\right] /D(q,\omega) \nonumber\\
&\chi_{\ua,\da}(q,\omega)&=\chi^0_{\ua,\ua}(q,\omega)\chi^0_{\da,\da}(q,\omega){\partial^2 e_{int} \over \partial n_\ua \partial n_\da} /D(q,\omega)
\label{chifinal}\enq 
where the denominator $D(q,\omega)$ is defined by
\beq
D(q,\omega)&=&\left[1-\chi^0_{\da,\da}(q,\omega){\partial^2 e_{int} \over \partial n_\da \partial n_\da}\right]\left[1-\chi^0_{\ua,\ua}(q,\omega){\partial^2 e_{int} \over \partial n_\ua \partial n_\ua}\right]\nonumber\\
&-&\chi^0_{\ua,\ua}(q,\omega)\chi^0_{\da,\da}(q,\omega)\left({\partial^2 e_{int} \over \partial n_\ua \partial n_\da}\right)^2
\label{D}
\enq
and, for simplicity, we have omitted the suffix $0$ in the density derivatives of $e_{int}$. The effects of the interaction is particularly crucial in the crossed $\chi_{\ua,\da}$ response. As we will explicitly discuss later the sign of the relevant interaction term $\partial^2 e_{int} / \partial n_\ua \partial n_\da$ is negative, reflecting the tendency of the system to pair opposite spins.
The poles of the response functions (zeros of $D(q,\omega)$) correspond to the undamped, discretized oscillations of the system (zero sound). They occur only in the presence of the interaction. Useful combinations of (\ref{chifinal}) are provided by the symmetric (s) and antisymmetric  (a) responses
$\chi_{s(a)}\equiv\chi_{\ua,\ua}+ \chi_{\da,\da}\pm 2\chi_{\ua,\da}$, also called density and spin responses. 

The concept of quasi-particle, and consequently result (\ref{chifinal}) for the response functions, applies to small wave vectors ($q \ll q_{F\sigma}$ where $q_{F\sigma}$ is the Fermi wavevector of the $\sigma$-specis). In this limit the  
free response  $\chi^0$, at zero temperature, reduces to the simple form:
\begin{equation}
\chi^0_{\sigma,\sigma^\prime}(q,\omega)= -{mq_{F\sigma}\over 2 \pi^2}g(\lambda_\sigma)\delta_{\sigma,\sigma^\prime}
\label{chig}
\end{equation}
with  $\lambda_\sigma=\omega/qv_{F\sigma}$ and $v_{F\sigma}\equiv \hbar q_{F\sigma}/m$. The function $g(\lambda)$ is defined by \cite{pines}
\beq
&g(\lambda)&= 1 - {\lambda\over 2}\ln{1+\lambda \over 1-\lambda}-i{\lambda \over 2}\pi  \; \; \; \; if \; \; \;   0<\lambda<1 \nonumber\\
&g(\lambda)&= 1 - {\lambda\over 2}\ln{\lambda +1 \over \lambda-1} \; \; \; \; \; \; \; \; \; \; \; \; \; \; if \; \; \;\lambda>1  \; .
\label{g}
\enq

From the knowledge of the response function one can  calculate the dynamic structure factor in each spin channel. At zero temperature the following relation holds for $\omega > 0$ 
\begin{equation}
S_\sigma(q,\omega) = -{1\over \pi} Im\chi_{\sigma,\sigma}(q,\omega) \; ,
\label{S}
\end{equation}
while $S_\sigma(q,\omega) = 0$ for $\omega <0$.
The function $S_\sigma(q,\omega)$ consists, in general, of a discretized peak (zero sound) and of a continuum of quasi-particle excitations. In the absence of interactions one finds $S^0_{\sigma}(q,\omega)= \lambda_\sigma m/2 \pi^2$ for $0<\lambda_\sigma<1$ and $0$ elsewhere. 

The dynamic structure factor obeys important sum rules \cite{pines,PS}. The most famous one is the {\it f}-sum rule $\int d\omega \omega S_{\sigma}(q,\omega)= n_\sigma q^2/2m$ which is exactly satisfied by the mean field results (\ref{chifinal}) for $\chi_{\sigma,\sigma}$, being directly related to the $1/\omega^2$ behavior of the response function at  large $\omega$. 
Another important sum rule is provided by the non-energy weighted moment, yielding the static structure factor
$\int d\omega  S_{\sigma}(q,\omega)= S_{\sigma}(q)$. This quantity is  directly related
to the Fourier tansform of the two-body correlation function. Landau's theory provides accurate information on the long wavelength (small $q$) behavior where $S_{\sigma}(q)$ is linear in $q$.

\begin{figure}[htb]
	\centering
		\includegraphics[height=5cm]{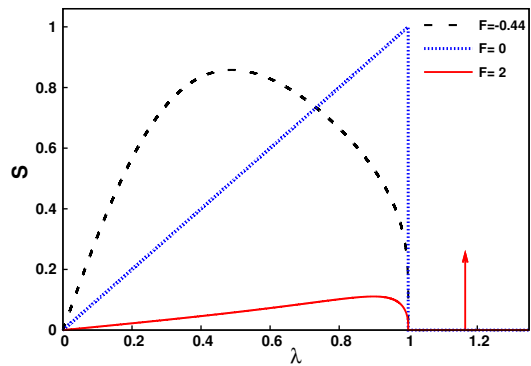}
	\caption{(Color online) Density (dashed, black) and spin (full line, red)  dynamic structure factor for an unpolarized Fermi gas in the normal phase at unitarity and $T=0$ as a function of  $\lambda=\omega/qv_F$. The red arrow  indicates the position of the spin zero sound solution. The ideal gas prediction is also shown (dotted, blue).}
	\label{fig:A}
\end{figure}

The simplest case is the symmetric configuration ($n_\ua=n_\da$). At very low temperature the corresponding ground state configuration is expected to be superfluid, the normal phase being available only at  higher temperatures where Landau's theory is not applicable. However, as already anticipated in the introduction, if one switches on  adiabatically the rotation of the confining trap, also in the unpolarized case the gas is predicted to phase separate at zero temperature into a superfluid core sorrounded by a normal shell \cite{ias}, so that the study of the dynamic behavior of the unpolarized normal gas might be  relevant for future experiments.  In this case the interaction coefficients determining  the response function  are conveniently written as $\partial^2 e_{int}/ \partial n_\ua \partial n_\ua=\partial^2 e_{int}/ \partial n_\da \partial n_\da=2/3 (\epsilon_F/n)(F_0^s+F_0^a)$ and $\partial^2 e_{int}/ \partial n_\ua \partial n_\da=2/3 (\epsilon_F/n)(F_0^s-F_0^a)$ where $F_0^{s(a)}$ are the  Landau parameters in terms of which   the density (s) and spin  (a) response functions take the familiar form \cite{pines}
\begin{equation}
\chi_{s(a)}= -{mq_F\over \pi^2} {g(\lambda)\over 1+F_0^{s(a)}g(\lambda)}  \; .
\label{chiunpol}
\end{equation}
The above Landau's  parameters are directly related to the compressibility $1/mc^2$ and  to the magnetic susceptibility $\chi_M$ of the gas according to 
\beq
&mc^2&=mc_0^2(1+F_0^s) \nonumber\\
&\chi_M&=\chi_{M0}/(1+F^a_0)
\label{chiequal}
\enq
where $c_0=v_F/\sqrt3$ and $\chi_{M0}$ are, respectively, the sound velocity and the magnetic susceptibility of the ideal gas.  
Since the gas is strongly interacting the evaluation of $F^s_0$ and $F^a_0$ requires a non perturbative  approach.  Monte Carlo calculations  of the equation of state and of the magnetic susceptibility in the normal phase yield the values $F_0^s= -0.44$ \cite{Carlson} and $F^a_0 \sim 2$ \cite{gprivate} at unitarity \cite{scattering}.  The behavior of the Landau's parameters of the unitary Fermi gas deeply differs from the one of liquid $^3He$ where $F_0^a$ is negative reflecting the  tendency of the system to behave like a ferromagnet. In our case $F_0^a$ is instead positive, reflecting the tendency of the system to pair opposite spins. An interesting consequence  is the appearence of a discretized pole in the spin response, associated with the propagation of spin zero sound, while the density response is Landau damped. In Fig.\ref{fig:A} we show the predicted values of the density and spin dynamic structure factor in the unpolarized case, using the values $F_0^s=-0.45$ and $F^a_0= 2.0$.   The spin zero sound solution takes place at the value $\omega=1.16qv_F$. The different behavior in the density and spin channels is evident  and shows up also in the low $q$ behavior of the static structure factors: $S_s(q) =1.26 S_0(q)$ and $S_a(q)=0.62 S_0(q)$ where $S_0(q)= 3/4 n q/q_F$ is the ideal Fermi gas prediction and $n$ is the gas density. The strength carried by the zero sound  is $\sim 80$ percent of the total spin strength. 

%\begin{figure}
%\begin{center}
%\includegraphics[width=0.6\columnwidth]{A.eps}
%\end{center}
%\caption{Density (dashed, black) and spin (full line, red)  dynamic structure factor for an unpolarized Fermi gas %in the normal phase at unitarity and $T=0$ as a function of  $\lambda=\omega/qv_F$. The red arrow  indicates the %position of the spin zero sound solution. The ideal gas prediction is also shown (dotted, blue). }
%\label{fig:Fig.1}
%\end{figure}

In the polarized case a first useful description of the interaction effects is provided by the low $x$ expansion of the equation of state. At unitarity the expansion takes the simple form $e_{int}= -(3/5)A \epsilon_{F\ua}n_\da$ with the value $A\sim 1$ predicted by both  Monte Carlo \cite{lobo,prok,seb} and diagrammatic calculations \cite{comb} and  where  $\epsilon_{F\ua}$ is the Fermi energy of the spin-up component. The expansion has been shown to provide an accurate description of the equation of state up to the largest   values of concentration achievable in the absence of rotation. In the small $x$ regime the relevant interaction parameter is $\partial^2 e_{int} / \partial n_\ua \partial n_\da=-2/5 A\epsilon_{F\uparrow}/n_\uparrow$.

\begin{figure}[htb]
	\centering
		\includegraphics[height=5cm]{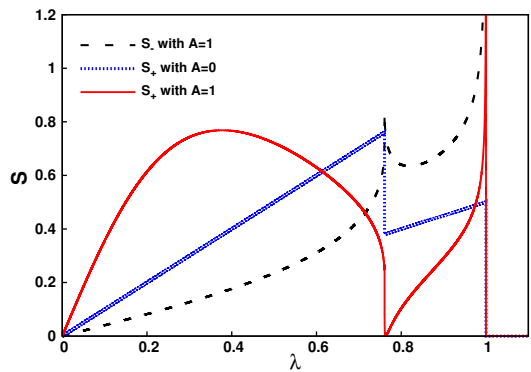}
	\caption{(Color online) Density (dashed, black) and spin (full line, red)  dynamic structure factor for a normal polarized ($x=n_\ua/n_\da= 0.44$) Fermi gas at unitarity and $T=0$ as a function of  $\lambda=\omega/qv_{F\ua}$. The ideal gas prediction is also shown (dotted, blue).}
	\label{fig:B}
\end{figure}

In Fig.\ref{fig:B} we show the predictions for  the density and spin dynamic structure factors calculated for a polarized Fermi gas at $x=0.44$. The comparison with the ideal gas prediction (dotted line) explicitly reveals the role of the interactions. Similarly to the unpolarized case interactions have opposite effects in the two channels. For the static structure factors, proportional to the integral of the dynamic structure factor, we find the results  $S_s(q) = 1.28 S_0(q)$ and $S_a(q)=0.83 S_0(q)$ where $S_0(q)=
3/4 (1+x^{2/3}) n_\uparrow q/q_{F\uparrow}=1.18 n_\uparrow q/q_{F\uparrow}$ is the ideal gas value. The cusps in the dynamic structure factors  at $\lambda_\ua=1$ and $\lambda_\ua=x^{1/3}$ reflect the existence of two Fermi surfaces in the polarized case.  In addition to the continuous structure  the response in the spin channel gives also rise to a discretized contribution (spin zero sound). This pole is however located too close to the continuum threshold $\lambda_\ua=1$ to be physically relevant, differently from what happens in the unpolarized case.

%\begin{figure}
%\begin{center}
%\includegraphics[width=0.6\columnwidth]{B.eps}
%\end{center}
%\caption{Density (dashed, black) and spin (full line, red)  dynamic structure factor for a normal polarized ($x=n_\ua/n_\da= 0.44$) Fermi gas at unitarity and $T=0$ as a function of  $\lambda=\omega/qv_{F\ua}$. The ideal gas prediction is also shown (dotted, blue). }
%\label{Fig.2}
%\end{figure}

The above results for the response function hold in the collisionless regime of the normal phase. When collisions become important  the dynamic behavior changes in a drastic way, being characterized by an ordinary (first) sound mode, giving rise to a sharp peak in the density response, and by a diffusive spin excitation. The first sound velocity is determined by the compressibility of the gas according to the thermodynamic relation $mc^2=n\partial \mu/\partial n$. For the unpolarized case we find $c=v_F(1+F^s_0)^{1/2}/\sqrt3$, corresponding to $\lambda = \omega/qv_F =0.43$, while for  the polarized gas one finds  $c=v_{F\ua}\left((1+x^{5/3}-Ax)/(1+x)\right)^{1/2}/\sqrt3$ \cite{compr}.  At $x=0.44$ the reduction of the sound velocity with respect to the ideal gas  ($A=0$) is still significant and comparable to the reduction calculated in the unpolarized case.  It is also useful to compare the above predictions with the superfluid behavior of the upolarized gas where  sound propagates hydrodynamically at the velocity $c=v_F(1+\beta)^{1/2}/\sqrt3$ with $\beta \sim =-0.58$. In the superfluid the spin excitations are instead gapped with the threshold given by $2\Delta$ where $\Delta$ is the single particle gap, of the order of the Fermi energy at unitarity. 

The transition from the collisionless to the collisional regime takes place when $\omega \tau \sim 1$ where $\tau$ is a typical collisional time. Using dimensional arguments one expects, at unitarity, the dependence $\hbar/\tau \propto \epsilon_F(k_BT/\epsilon_F)^2$, where the $T^2$ factor originates from the Pauli principle.  The value of the  coefficient of proportionality is estimated to be  $\sim 4$ both in the unpolarized case \cite{bruunprivate} and in the highly polarized limit \cite{bruun} (in the latter case  $\epsilon_F$ is the Fermi energy of the majority component). The conditions for applying Landau's theory in the collisionless regime are not consequently too severe, requiring frequencies satisfying the condition:
\begin{equation}
\epsilon_F \left({k_BT\over \epsilon_F}\right)^2 \ll \hbar \omega \ll \epsilon_F \; .
\label{condition}
\end{equation}
The condition $\hbar \omega \gg \epsilon_F \left(k_BT/ \epsilon_F\right)^2$ is in particular much less severe than in the case of the spin dipole oscillation of the trapped gas which takes place at frequencies of the order of the harmonic oscillator frequency \cite{lobo}. 

In conclusion we have shown that at unitarity the response function of a Fermi gas in its normal phase is sizably affected by the interactions and  behaves quite differently in the density and spin channels. Both mean field and collisional effects are predicted to take place in ranges of temperatures and frequencies of reasonably easy access
in two photon Bragg spectroscopy experiments.

Useful discussions with G. Bruun, S. Giorgini, S. Pilati and M. Zwierlein are acknowledged. This work was supported by MIUR and by the Euroquam Fermix programme. The kind hospitality at the Center for Ultracold Atoms in Cambridge (US), where part of this work was carried out, is also acknowledged.


\begin{thebibliography}{99}
\bibitem{scattering} The MC values for the Landau parameters  $F_0^s$ and $F_0^a$ almost exhaust the forward scattering amplitude sum rule \cite{pines} $\sum_\ell \left[F_\ell^s/(1+F^s/(2\ell+1))+F_\ell^a/(1+F^a/(2\ell+1))\right]= 0$, thereby confirming that   the parameters $F_\ell$ with $\ell \ge 1$ should not play a major role at unitarity.
\end{thebibliography}
\end{document}